\documentclass[conference]{IEEEtran}
\def\IEEEtitletopspace{18pt}
\IEEEoverridecommandlockouts

\usepackage{fancyhdr}
\fancypagestyle{firstpage}{%
  \lhead{This work has been accepted for publication in the IEEE ICC 2025 Conference.}
}

 \usepackage{color}
 \usepackage{tabularray}
\usepackage[table,xcdraw]{xcolor}
\usepackage{dblfloatfix}
\usepackage{cite}
\usepackage{multicol}
\usepackage{diagbox}
\usepackage{tabularray}
\usepackage{amssymb}
\usepackage{amsfonts}
\usepackage{amsmath}
\usepackage[T1]{fontenc}
\usepackage[utf8]{inputenc}
\usepackage[english]{babel}
\usepackage{pifont}
\usepackage{booktabs}
\usepackage{amsmath}
\usepackage{mathtools}
\usepackage{ellipsis}
\usepackage{textcomp}
\usepackage{cite}
\usepackage[pdftex]{graphicx}
\usepackage{subfigure}
\usepackage{epsfig}
\usepackage{epstopdf}
\usepackage{caption}
\usepackage{graphicx}
\usepackage{textcomp} 
\definecolor{Mycolor1}{HTML}{ff0000}
\definecolor{Mycolor2}{HTML}{13beb8}
\definecolor{Mycolor3}{HTML}{ffea00}
\definecolor{Mycolor4}{HTML}{4dd0e1}
\definecolor{Mycolor5}{HTML}{ab47bc}
\definecolor{Mycolor6}{HTML}{ffa726}
\definecolor{Mycolor7}{HTML}{bcaaa4}
\definecolor{Mycolor8}{HTML}{43a047}
\definecolor{Mycolor9}{HTML}{81d4fa}
\definecolor{Mycolor10}{HTML}{999900}
\definecolor{Mycolor11}{HTML}{FFEBCC}
\definecolor{Mycolor12}{HTML}{D1E9E2}
\usepackage{algorithm}
\definecolor{customblue}{rgb}{0.30, 0.65, 0.89} 
\definecolor{customorange}{rgb}{1.00, 0.65, 0.20} 
\usepackage{fixltx2e}
\usepackage{algpseudocode}
\usepackage{epstopdf}
\usepackage{psfrag}
\usepackage{grffile}
\usepackage{amsthm}

\usepackage{caption}

\theoremstyle{definition}

\theoremstyle{remark}

\theoremstyle{plain}

\newcommand{\RNum}[1]{\uppercase\expandafter{\romannumeral #1\relax}}
\usepackage{setspace}

\def\BibTeX{{\rm B\kern-.05em{\sc i\kern-.025em b}\kern-.08em
    T\kern-.1667em\lower.7ex\hbox{E}\kern-.125emX}}
\begin{document}
\title{Environment-Aware Scheduling of URLLC and Sensing Services for Smart Industries}
\author{\IEEEauthorblockN{ Navid Keshtiarast, Pradyumna Kumar Bishoyi, and Marina Petrova}
\IEEEauthorblockA{Mobile Communications and Computing, RWTH Aachen University, Aachen, Germany \\
Email: \{navid.keshtiarast, pradyumna.bishoyi, petrova\}@mcc.rwth-aachen.de}
}
\maketitle
\thispagestyle{firstpage}
\begin{abstract}
 In this paper, we address the problem of scheduling sensing and communication functionality in an integrated sensing and communication (ISAC) enabled base station (BS) operating in an indoor factory (InF) environment. The BS is performing the task of detecting an AGV while managing downlink transmission of ultra-reliable low-latency communication (URLLC) data in a time-sharing manner. Scheduling fixed time slots for both sensing and communication is inefficient for the InF environment, as the instantaneous environmental changes necessitate a higher frequency of sensing operations to accurately detect the AGV. To address this issue, we propose an environment-aware scheduling scheme, in which we first formulate an optimization problem to maximize the probability of detection of AGV while considering the survival time constraint of URLLC data. Subsequently, utilizing the Nash bargaining theory, we propose an adaptive time-sharing scheme that assigns sensing duration in accordance with the environmental clutter density and distributes time to URLLC depending on the incoming traffic rate. Using our own Python-based discrete-event link-level simulator, we demonstrate the effectiveness of our proposed scheme over the baseline scheme in terms of probability of detection and downlink latency.
\end{abstract}
%
\section{Introduction}
Automated guided vehicles (AGVs) will lie at the heart of tomorrow's smart factory scenario, executing a variety of logistical tasks autonomously and efficiently with minimal direct human involvement \cite{Rodrigo_mag}. These AGVs usually move around the halls in a predefined track and follow the instructions provided by the central unit \cite{factory_positioning2023}. The simultaneous presence of AGVs and human workers at the indoor factory (InF) environment poses a new safety challenge \cite{3GPP3}. In order to ensure the rigorous safety operation of AGVs, it is essential to have a reliable infrastructure that can accurately detect the position of AGVs on the track and provide precise awareness of the surrounding environment, including obstacles and humans \cite{MoeZin_icassp}. In this respect, integrated sensing and communication (ISAC) systems appear as an ideal platform for InF environments, given their capability to identify targets of interest while simultaneously delivering communication services \cite{Silvio_servey}. 
\begin{figure}[t]
\centering
\includegraphics[width=0.45\textwidth,trim = 0mm 95mm 0mm 10mm,clip]{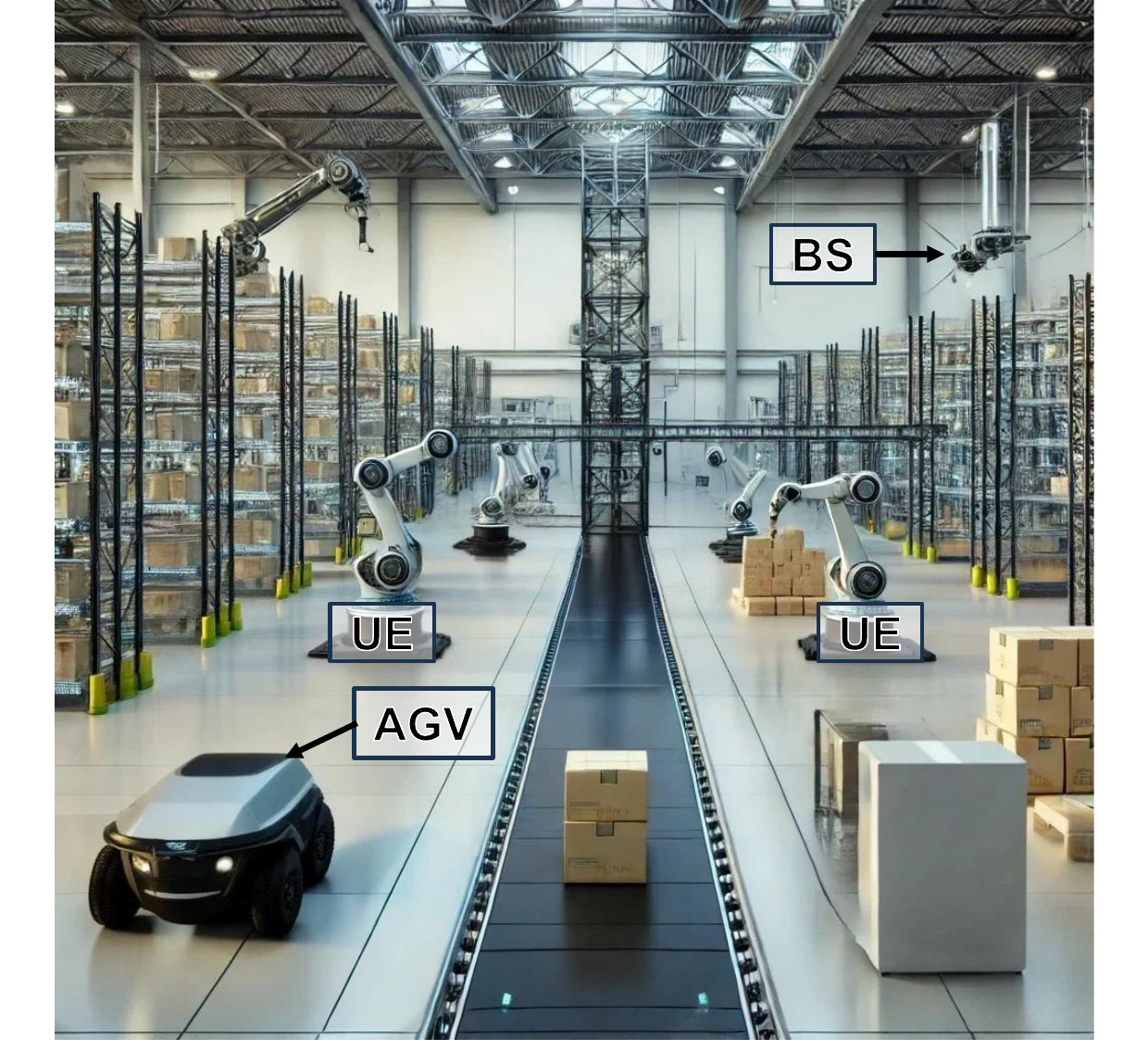}
	\caption{Illustration of smart InF environment\protect\footnotemark.}
	\label{fig:system_fig}
 \vspace{-8mm}
\end{figure}
\footnotetext{Figure generated using DALL-E 3}

The next-generation base stations (BSs) with ISAC capability are expected to play an important role in smart industries by providing ultra-reliable low-latency communication (URLLC) communication services as well as tracking the trajectory of AGVs, as shown in Fig. \ref{fig:system_fig}. However, scheduling resources between sensing and communication functionality at the BS is quite challenging in the InF environment due to the following factors: (i) random mobility of workers and AGVs, (ii) varying propagation conditions due to clutter density, (iii) inter-BS interference, and (iv) diverse and stringent service requirements of communication and sensing function. All these factors introduce uncertainty that must be carefully considered for resource scheduling \cite{MoeZin_icassp}. For instance, when there are more workers adjacent to the track or the AGV traverses through a cluttered environment, the BS must allocate additional time for sensing to detect unexpected events on the track. Conversely, in a less cluttered environment, the BS can allocate more resources to URLLC data communication. Moreover, in contrast to conventional communication-centric scheduling, ISAC-based scheduling must be adaptable to the external environment, as the efficacy of sensing functions (whether detection or localization) is significantly influenced by the surrounding conditions \cite{Silvio_servey,navid_lnet2024}. Consequently, the primary focus of this work is to design an adaptive environment-aware resource allocation algorithm at ISAC-enabled base stations operating in the InF environment. 

A few recent works discuss the coexistence of sensing and URLLC communication. In \cite{ding_mobicom22}, a joint precoding scheme is proposed to improve the radar performance while considering the QoS of URLLC transmission. In \cite{vacarubio2023}, the authors proposed a deep learning-based framework to allocate the time slot between sensing and communication for mmWave ISAC system. Authors in \cite{Zinat_wcnc2024} consider a cell-free massive MIMO system and propose a power allocation scheme to jointly improve the multi-target detection and URLLC communication performance. All of these above works focus on the sensing in outdoor scenarios and do not consider the complexity arising from the dynamic InF environment. 
In \cite{ramos2024}, the authors propose a predictive beamforming scheme for the InF environment by incorporating the sensing information obtained from the beam training phase and utilizing the user trajectory information. The framework solely focuses on sensing without considering the analysis of communication performance. To this end, the issue of adaptive resource allocation between sensing and URLLC communication, considering the dynamic and uncertainty in the environment and the traffic, remains unanswered.

In this paper, we study an ISAC system for the InF environment, where a BS communicates with URLLC user equipment while simultaneously sensing a moving AGV in a cluttered environment. Similar to \cite{navid_wcnc2024}, we consider that the sensing and communication functionality at BS is time-multiplexed and allocated in time cycles. In each time cycle, the BS decides how much it should allot for sensing and communication based on its conflicting requirements. Thus, we formulate an optimization problem that maximizes the probability of detection of AGV, subject to a delay deadline constraint for URLLC traffic, while taking the environmental clutter density into account. Further, we solve the problem by modelling the time-sharing between sensing and communication as a bargaining process, and employ the well-known \textit{Nash bargaining theory}, which yields fair and Pareto-efficient outcome \cite{nash1950bargaining,touti_globe,broadband_bargain,bishoyi2024}. Unlike traditional methods like proportional fairness, which equally weights all services, and round-robin, which sequentially allocates resources without priority adjustments, Nash bargaining solution (NBS) can dynamically shift resources to prioritize objectives based on service demands. For instance, when URLLC traffic is high, NBS can favor communication over sensing resources, or vice versa, hence self-inforcing in our scenario. The main contributions of this work are summarized as follows:
\begin{itemize}
\item We propose a novel time-allocation solution that dynamically adjusts the time dedicated to sensing and communication functionality in an InF environment. By employing the Nash Bargaining Solution, we optimize the allocation to maximize the probability of detecting the AGV while maintaining the stringent QoS requirement for URLLC traffic. This leads to a fair and efficient resource allocation that enhances sensing performance without adversely affecting URLLC traffic. 
\item To validate the effectiveness of our proposed scheme, we developed a discrete-event link-level simulator using the Python Simpy library. We model the dynamic environment of the industry, such as the dynamic movement of the AGV, variable URLLC traffic loads and complex propagation conditions, which are given in the 3GPP TR 38.901 channel model\cite{3GPP2}. Our simulation results demonstrate that our adaptive scheme outperforms the baseline and effectively balances the dual objectives of sensing and communication.
\end{itemize}

The remainder of this paper is organized as follows: Section II provides a detailed discussion of the system model and the performance metrics. In Section III, we present our proposed approach for resource allocation.  Section IV offers a comprehensive analysis of the simulation results and discusses their implications. Finally, Section V concludes the paper.
\section{System Model}\label{System_Model}
We consider an indoor factory scenario in which an ISAC-enabled 5G new radio (NR) BS is deployed to establish downlink communication with $I$ industrial machines and also to sense an automated guided vehicle (AGV). The BS is equipped with $N_{tx}$ antennas deployed as a horizontal uniform linear array (ULA) with half-length separation and uses orthogonal frequency division multiplexing (OFDM) signal for both sensing and communication. We assume that the BS is performing sensing and communication functions in a time-sharing manner to achieve isolation between two functionalities. In the case of sensing, the BS is performing monostatic sensing. The AGV is moving through the industrial hall for goods storage with a predetermined reference track. The objective of the BS is to sense and detect if the AGV is present on the track or not. In case of sensing, we consider a set of industrial machines denoted as $ \mathcal{I}=\{1,2,\cdots,I\} $ and connected to the BS. We consider only downlink traffic, where the BS transmits the URLLC traffic to the machines.

In this work, the smallest unit of time considered for the analysis is one 5G subframe of duration $1$ ms. The total number of subcarriers and symbols available for both sensing and communication within one subframe is denoted as $N_{sc}^{tot}$ and $M_{sym}^{tot}$, respectively. In the subsequent subsections, we first explain the 3GPP-specified InF channel model and then explain the sensing and communication operation in detail.
\subsection{Channel Model}
InF environments are characterized by unique propagation conditions due to obstructions from machinery, high ceilings, and the presence of metallic objects, which introduce significant variability in signal behaviour. To model these channel conditions accurately, we adopt the 3GPP TR 38.901 \cite{3GPP2} channel model, which incorporates clutter density and antenna heights for Line of Sight (LoS) probability estimations. Further, it specifies path loss models for both LoS and Non-Line of Sight (NLoS) conditions.
The LoS path loss equation, which is applicable to all scenarios, is defined as follows:

\begin{equation}
\begin{aligned}
PL_{\text{LOS}}(f_c, d_{3D})& = 31.84 + 21.50 \log_{10}(d_{3D}) \\& + 19.00 \log_{10}(f_c), \quad \sigma = 4.3,
\end{aligned}
\end{equation}
where \(f_c\) is the frequency of operation (\(f_c\)), \(d_{3D}\) is the three-dimensional distance between the transmitter and the receiver, and \(\sigma\) is the standard deviation that models shadow fading depending on the scenario.

For NLoS conditions with high base station height, the path loss equations vary depending on the specific scenario:

\begin{itemize}

\item Sparse clutter with high BS height (SH):
\begin{align}\label{lowbs_sl}
&PL_{\text{SH}}(f_c, d_{3D}) = 32.4 + 23.0 \log_{10}(d_{3D}) + 20 \log_{10}(f_c) \nonumber\\
&PL_{\text{NLOS}} = \max(P_{\text{LSH}}, P_{\text{LLOS}}), \quad \sigma = 5.9 
\end{align}
\item Dense clutter with high BS height (DH):
\begin{align}\label{nloslowbs_sl}
&PL_{\text{DH}}(f_c, d_{3D}) = 33.63 + 21.9 \log_{10}(d_{3D}) + 20 \log_{10}(f_c) \nonumber\\
&PL_{\text{NLOS}} = \max(P_{\text{LDH}}, P_{\text{LLOS}}), \quad \sigma = 4.0 
\end{align}
\end{itemize}

In addition, the probability of transitioning from NLoS to LoS as the distance increases can be described by an exponential decay function, i.e.,  
\begin{equation}
P_{\text{LoS,sec}}(d_{2D}) = \exp\left(-\frac{d_{\text{2d}}}{r_{\text{sec}}}\right).
\end{equation}
The variable \( r_{\text{sec}} \) is calculated differently depending on the scenario and can be expressed as:
\[
r_{\text{sec}} = 
\begin{cases} 
\frac{d_{\text{clutter}}}{\ln(1-r)} & \text{for the SL and DL scenarios}, \\
\frac{d_{\text{clutter}}}{\ln(1-r)}\frac{ h_{\text{BS}} - h_{\text{UT}}}{h_{c}-h_{\text{UT}}} & \text{for the SH and DH scenarios.}
\end{cases}
\]
In these equations, the clutter density \( r \) has a value that ranges between 0 and 1. Other key parameters include \( d_{\text{clutter}} \) (the size of the clutter), \( h_{\text{c}} \) (the height of the clutter), \( h_{\text{UT}} \) (the height of the User Terminal antenna), and \( h_{\text{BS}} \) (the height of the base station antenna). Lastly, \( d_{2D} \) represents the two-dimensional distance between the base station and the user equipment.
\vspace{-0.1cm}
\subsection{AGV Detection}
The BS transmits a sensing signal with $M_{sym}^{s}$ OFDM symbols and $N_{sc}^{s}$ active subcarriers. The complex baseband representation of transmitted sensing signal $s_m (t)$ is
\begin{equation}\label{tx_sens_signal}
    s_m(t) = \sum_{n =0}^{N_{sc}^{s}-1} \sqrt{P_{n,m}}x_{n,m}e^{j2\pi n\Delta ft}\text{rect}\bigg(\frac{t-mT_{sym}}{T_{sym}}\bigg),
\end{equation}
where $x_{n,m}$ is the modulation symbol transmitted in the $n$-th subcarrier of the $m$-th symbol and $P_{n,m}$ is the corresponding allocated power. $\Delta f$ denotes the subcarrier spacing. $T_{sym} = T+T_{CP}$ is the total symbol duration, $T_{CP}$ is the cyclic prefix duration, and $T=1/\Delta f$ is the OFDM symbol duration. 

With full-duplex capability assumption at the BS transceiver, the reflected signal from the AGV is captured at the BS. Further, we assume AGV as a point target, and its relative velocity is $v$. The BS receives the time domain echo signal and transforms it into the frequency domain via fast Fourier transform (FFT) for further radar processing. The received symbol across the subcarrier $n$ and symbol $m$ after FFT processing is
\begin{equation}\label{rx_sens_signal}
    Y(n,m) = \alpha_{n,m}x_{n,m}e^{(j2\pi f_d mT_{sym}-j2\pi n\Delta f\tau)} + z_{n,m},
\end{equation}
where $\alpha_{n,m}$ is the complex channel gain for subcarrier $n$ and symbol $m$, $f_d = \frac{2vf_c}{c}$ is the doppler shift, and $z_{n,m}$ is the noise component.
\subsubsection{Resource scheduling for sensing}
The BS detects the AGV by collecting and processing the echoes reflected from it. The expression for received echo power is 
\begin{equation}
    P_r = \frac{P_{tx}G^2f_c^24\pi\sigma_{rcs}}{{PL}^2c^2},
\end{equation}
where $P_{tx}$ is the transmitting power $G = 4\pi/\phi^2$ is the antenna gain over the main beam of width $\phi = [0,2\pi]$, $\sigma_{rcs}$ is the radar cross section (RCS) of the AGV, $PL$ is the path loss (based on Eq. \eqref{lowbs_sl}-\eqref{nloslowbs_sl}), $c$ is the speed of the light, $f_c$ is the operating frequency. Considering thermal noise at the BS receiver, the noise power can be written as $P_n = k_B T_{therm}FN_{sc}^{s}\Delta f$, where $k_B$ is Boltzman constant, $T_{therm}$ is the standard room temperature (in Kelvin), and $F$ is the noise figure. The corresponding signal-to-noise ratio (SNR) on each OFDM symbol is 
\begin{equation}
    \gamma =  \frac{P_r}{P_n} = \frac{P_{tx}G^2f_c^24\pi\sigma_{rcs}}{{PL}^2c^2k_B T_{therm}FN_{sc}^{s}\Delta f}.
\end{equation}
The SNR is increased due to the multiplicative gain given by the number of subcarriers $N_{sc}^{s}$ and OFDM symbols $M_{sym}^{s}$, resulting in an SNR gain of
\begin{equation}\label{gamma_pd_eq}
    \gamma_{sens} =  \gamma N_{sc}^{s}M_{sym}^{s} = \frac{P_{tx}G^2f_c^24\pi\sigma_{rcs} M_{sym}^{s}}{{PL}^2c^2k_B T_{therm}F\Delta f}.
\end{equation}
The probability of detection is given by \cite{Silvio_servey}
\begin{equation}\label{radar_pd_eq}
    P_d = \mathcal{Q}_1\bigg(\sqrt{\gamma_{sens}(M_{sym}^{s})}, \sqrt{2P_{fa}}\bigg), 
\end{equation}
where $\mathcal{Q}_1(\cdot)$ is the Marcum Q-function of order $1$ and $P_{fa}$ is the probability of false alarm. The value of $P_d$ is dependent on the coherent processing time, i.e. the actual gain in SNR is obtained from the observations of $M_{sym}^{s}$ OFDM symbols. Further, given the false alarm rate and probability of detection, we can obtain the target SNR requirement $\gamma_{sens}^{min}$ that is necessary to achieve robust target detection performance. Therefore, to achieve the $\gamma_{sens}^{min}$ when the AGV is at a distance $d$, we can derive the minimum number of OFDM symbols required for the coherent processing (from Eq. \eqref{gamma_pd_eq}), i.e.,
\begin{equation}\label{min_req_sens}
    M_{req}^{s} \geq \frac{\gamma_{sens}^{min}{PL}^2c^2k_B T_{therm}F\Delta f}{P_{tx}G^2f_c^24\pi\sigma_{rcs}}.
\end{equation}
Clearly, we observe that in case of a more cluttered environment, the path loss ($PL$) will be higher, and the BS has to assign more symbols for sensing. Therefore, to guarantee the target detection, the BS time-allocation for sensing must satisfy,
\begin{equation}\label{sens_min}
    T_{sens}\geq M_{req}^{s}T_{sym}.
\end{equation}
\subsection{Resource scheduling for URLLC traffic}
The end-to-end delay and overall loss probability of each packet characterize the QoS requirement of URLLC traffic \cite{Salah_jsac}. We have considered downlink traffic in our scenario. Therefore, the QoS requirement in our case will be the BS transmits a packet of size $p$ successfully within an end-to-end delay limit of $D_{max}$ with a probability of failure at most $\sigma_{urllc}$. However, the downlink transmission is affected by the channel condition. To support the high reliability over fading wireless channels in 5G NR, different Hybrid Automatic Repeat reQuest (HARQ) retransmission mechanisms are proposed. Further, to ensure successful delivery, the delay caused by the retransmissions must be considered, i.e. the total time elapsed from the packet generation to successful delivery of packet with retransmission(s) must be within the delay limit $D_{max}$. This introduces a constraint on the time resources allocated for the URLLC traffic. 

For URLLC traffic, we consider the resource block (RB) to be the smallest resource unit assigned to the transmission of downlink data. We assume that the BS has the channel quality index (CQI) information of the downlink channel before the transmission of each packet, and a robust modulation and coding scheme (MCS) is assigned to ensure a lower block error rate (BLER). The total number of RBs scheduled for transmitting a packet size of $p$ (in bits) is
\begin{equation}
    R = \bigg\lceil \frac{p}{\eta W\kappa T_{sym}} \bigg\rceil,
\end{equation}
where $\eta$ is the MCS of unit bits/s/Hz, $w$ is the bandwidth per RB, $\kappa$ is a non-negative integer corresponding to a number of OFDM symbols, and $\kappa T_{sym}$ is the TTI duration. 

Let $\delta$ be the BLER for each downlink transmission, which is affected by both the channel condition and the choice of MSC \cite{3GPP1}. If the BLER value $\delta$ is higher than the reliability constraint $\sigma_{urllc}$, then it may take several retransmissions to achieve the requirement. Considering each transmission/retransmission as an independent event, the expression for the probability of failure is \cite{Salah_jsac}
\begin{equation}\label{fail}
    \mathbb{P}_{fail} = (\delta)^{n_{tx}}\leq \sigma_{urllc},
\end{equation}
where $n_{tx}= 1 + n_{retx}$ is the total number of transmissions and $n_{retx}$ is the total number of retransmissions. From Eq. \eqref{fail}, for a given BLER and reliability constraint, we can obtain the total number of retransmissions as
\begin{equation}\label{faill}
    n_{tx} = \bigg\lceil \frac{\log(\sigma_{urllc})}{\log(\delta)} \bigg\rceil.
\end{equation}
Therefore, given delay limit $D_{max}$ and maximum possible retransmissions $n_{retx}$, the minimum time that BS has to reserve for URLLC communication is
\begin{equation}\label{comm_min}
    M_{req}^{c}= \min\bigg(n_{tx}T_{sym}, D_{max}\bigg).
\end{equation}
\subsection{Problem Formulation}
We are interested in determining the optimal time allocation between sensing and communication in such a way that the AGV can be detected with higher probability and the delay requirement of URLLC traffic is maintained. From Eq. \eqref{radar_pd_eq}, we observe that the AGV's Probability of detection ($P_d$) increases when we assign more OFDM symbols ($N_{sym}^s$) for sensing. However, such allocation of symbols will minimize the communication duration, thereby severely degrading the QoS of URLLC traffic. Furthermore, a fixed time allocation approach is not appropriate since the environment (clutter density) in the industrial scenario is dynamic. Therefore, our objective is to develop an environment-aware adaptive time-allocation scheme that can maximize the probability of detection of the AGV while maintaining the QoS of URLLC. The corresponding optimization problem is formulated as follows:
\begin{alignat}{2}
& \underset{N_{sym}^s,N_{sym}^c}{\max} \hspace*{0.5cm}  P_d(\gamma_{sens}) \label{objective} \\
& \text{s.t.}\hspace*{0.5cm}\eqref{sens_min}, \eqref{comm_min} \nonumber \\
&\hspace*{1cm}   D_i^{urllc}(N_{sc}^c,N_{sym}^c) \leq D_i^{serv},\hspace*{0.2cm}\forall i \in \mathcal{I}\label{com_con2}\\
&\hspace*{1cm} N_{sym}^s + N_{sym}^c \leq N_{sym}^{tot}. \label{sym_con4}
\end{alignat}  
The Eq. \eqref{com_con2} ensures that the delay for all URLLC traffic remains within its survival time ($D_i^{serv}$). Meanwhile, the Eq. \eqref{sym_con4} ensures that the total OFDM allocation does not exceed $N_{sym}^{tot}$.

\section{Nash Bargaining-based Optimal Time allocation}
We employ the Nash bargaining solution (NBS) \cite{nash1950bargaining} to optimize the allocation of time between sensing and communication, thereby balancing the dual objectives of AGV detection and downlink communication of URLLC traffic. NBS has been widely used in different domains as a fair and efficient solution method to achieve a good tradeoff between the individual performance objective and the global objective \cite{broadband_bargain,bishoyi2024}. In our scenario, we want to improve the sensing performance without detrimental performance on URLLC traffic. We can model it as a 2-player bargaining game where sensing and communication functions can be considered two players. Let $\mathcal{U}_{sens}$ and $\mathcal{U}_{com}$ denote the utility function for sensing and communication, respectively. 

Let $n_{sym}^sT_{sym}$ be the total time allotted for sensing. Then, the utility gain for the sensing is defined as the additional time that is allocated for sensing beyond the minimum requirement $M_{req}^{s}$ (from Eq. \eqref{min_req_sens}), i.e.,
\begin{equation}\label{sens_uti}
    \mathcal{U}_{sens} = (n_{sym}^s - M_{req}^{s})T_{sym}.
\end{equation}
Similarly, the utility gain for the URLLC communication is
\begin{equation}\label{comm_uti}
    \mathcal{U}_{com} = (n_{sym}^c - M_{req}^{c})T_{sym}.
\end{equation}
\begin{figure}[t]
\centering
\includegraphics[width=0.45\textwidth,trim = 95mm 61mm 105mm 29mm,clip]{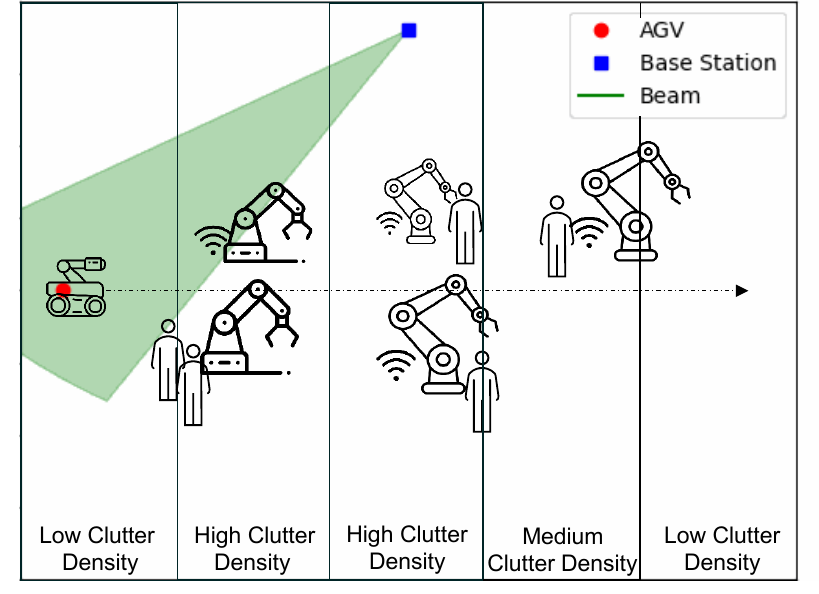}
\vspace{-1mm}
	\caption{System concept.}
	\label{fig:System_model1}
 \vspace{-7mm}
\end{figure}
Clearly, $M_{req}^{s}$ and $M_{req}^{c}$ can be envisioned as disagreement points in the bargaining, and their values change according to the environmental requirement. Now, we define the NBS-based utility maximization problem as follows:
\begin{alignat}{2}
& \underset{n_{sym}^s,n_{sym}^c}{\max} \hspace*{0.5cm} \mathcal{U}^{NBS} = \mathcal{U}_{sens}\mathcal{U}_{com}\label{objective1} \\
& \text{s.t.}
\hspace*{1cm}  \mathcal{U}_{sens} \geq 0, \quad \mathcal{U}_{com} \geq 0 .\label{sym_con4_1}
\end{alignat} 
Since solving the above NBS problem is quite challenging, we convert the product form of objective function in \eqref{objective1} into a more tractable sum of logarithmic terms, i.e.
\begin{align}\label{nbs_mid}
\mathcal{\widetilde{U}}^{NBS}(n_{sym}^s,n_{sym}^c) &= \log(\mathcal{U}_{sens}) + \log(\mathcal{U}_{com}).
\end{align}

Let $(n_{sym}^{s*},n_{sym}^{c*})$ denote the optimal NBS solution which maximizes the value $\mathcal{\widetilde{U}}^{NBS}$ and provides optimal solution for problem stated in \eqref{objective}. Setting the first-order derivative of Eq. \eqref{nbs_mid} with respect to $n_{sym}^{s}$ and setting that to zero, we can obtain the expression of the optimal solution of $n_{sym}^{s*}$
\begin{equation}\label{nbs_sens_fin}
\begin{multlined}[t]
    n_{sym}^{s*} = \frac{1}{2} \Bigg[ \frac{\gamma_{sens}^{min}{PL}^2c^2k_B T_{therm}F\Delta f}{P_{tx}G^2f_c^2\sigma_{rcs}4\pi} + N_{sym}^{tot} -  M_{req}^{c}\Bigg]\nonumber.
\end{multlined}
\end{equation}
Since the total available OFDM symbols is $N_{sym}^{tot}$, subtracting the $n_{sym}^{s*}$ number of symbols allotted for sensing, we obtain the optimal value for $n_{sym}^{c*}$ as,  
\begin{equation}
    n_{sym}^{c*} = N_{sym}^{tot} - n_{sym}^{s*}.
\end{equation}
The total time allotted for communication ($n_{sym}^{c*}$) is shared equally among all the URLLC traffic users.

\section{Performance Evaluation}
In this section, we evaluate the performance of the proposed resource allocation approach within a simulated InF environment. The environment is modeled as a $200~m\times200~m$ industrial hall divided into multiple sectors, each characterized by different clutter densities. A single 5G NR BS is centrally positioned within the hall, serving multiple URLLC user equipments and equipped with the capability to detect an AGV in sensing mode as it moves across the sectors as shown in Figure~\ref{fig:System_model1}. We have developed a time-discrete link-level simulator using the SimPy library for our simulation and evaluation. This simulator accurately models various critical aspects of the industrial environment, including network traffic behavior, mobility patterns of both user equipment and the AGV, and varying clutter densities.

The cumulative load of URLLC traffic at the base station is modelled as a Poisson process with a packet arrival rate of $\lambda_{\text{URLLC}}$ packets per second. The URLLC traffic rate varies randomly every second, simulating fluctuations between low and high traffic demands generated by UEs distributed throughout the environment. The BS leverages physical layer reports of channel conditions to adaptively select the MCS from CQI Table 2 from 3GPP TS 38.214\cite{3GPP1}. We set a fixed BLER target of $0.001$ for URLLC users. The retransmission limit is set to $3$ if the packet transmission fails to meet the required BLER \cite{3GPP2}. At the start of the simulation, the AGV is placed at a random location and it randomly moves either left or right, with velocity is set to $4$ m/s. We compare our resource allocation approach with the Round-Robin-based scheduling between sensing and communication. All simulation parameters are presented in Table \ref{sec:ScenariosandSetup}.
\begin{figure}[t]
\centering
\includegraphics[width=0.45\textwidth,trim = 0mm 0mm 0mm 0mm,clip]{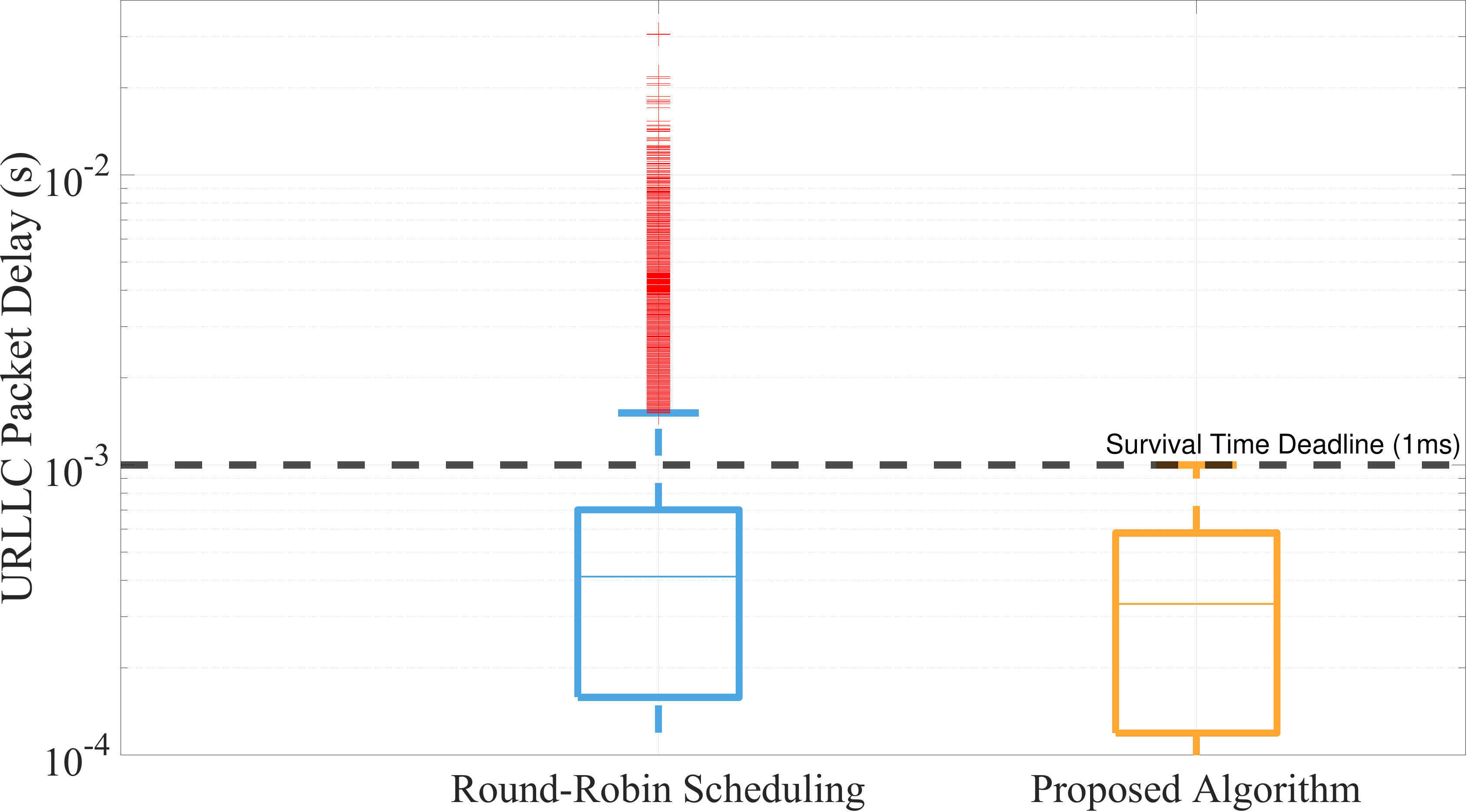}
\vspace{-1mm}
	\caption{Distribution of URLLC packet delay.}
	\label{fig:results_1}
 \vspace{-7mm}
\end{figure}
\begin{table}[h!]
\centering
\vspace{-2mm}
\caption{Simulation Parameters}
\begin{tabular}{|c|c|}
\hline
\textbf{Parameters} & \textbf{Value} \\
\hline
Frequency & 3.5 GHz \\
\hline
Total Bandwidth, SCS & 100 MHz, 30 kHz \\
\hline
Beamwidth & 27° \\
\hline
Transmitting Power & 30 dBm \\
\hline
RCS & 7 dBsm \\
\hline
No. of URLLC users & 10 \\
\hline
$\lambda_{URLLC}$  & $[1000, 2000, \cdots, 400000]$ \\
\hline
Survival time & 1 ms \\
\hline
Channel model & InF from 3GPP Eq. \eqref{lowbs_sl}-\eqref{nloslowbs_sl} \\
\hline
Simulation duration & 20 s \\
\hline
\end{tabular}
\label{sec:ScenariosandSetup}
\vspace{-2mm}
\end{table}

Figure \ref{fig:results_1}  illustrates the distribution of the end-to-end URLLC packet delays achieved under the two different schedulers.   The URLLC traffic rate is varied every second for each of the ten users simulating dynamic load conditions. The results clearly show that despite the fluctuating traffic, the proposed algorithm based on Nash Bargaining consistently outperforms the Round-Robin scheduling approach, keeping the URRLC packet delay below 1 ms, thereby ensuring that the survival time is never exceeded. The Round-Robin algorithm, on the other hand, produces numerous outliers, with some URLLC packet delays approaching 0.1 seconds, which are far too high to meet the stringent URLLC requirements. 

\begin{figure}[t]
\centering
\includegraphics[width=0.45\textwidth,trim = 0mm 0mm 0mm 0mm,clip]{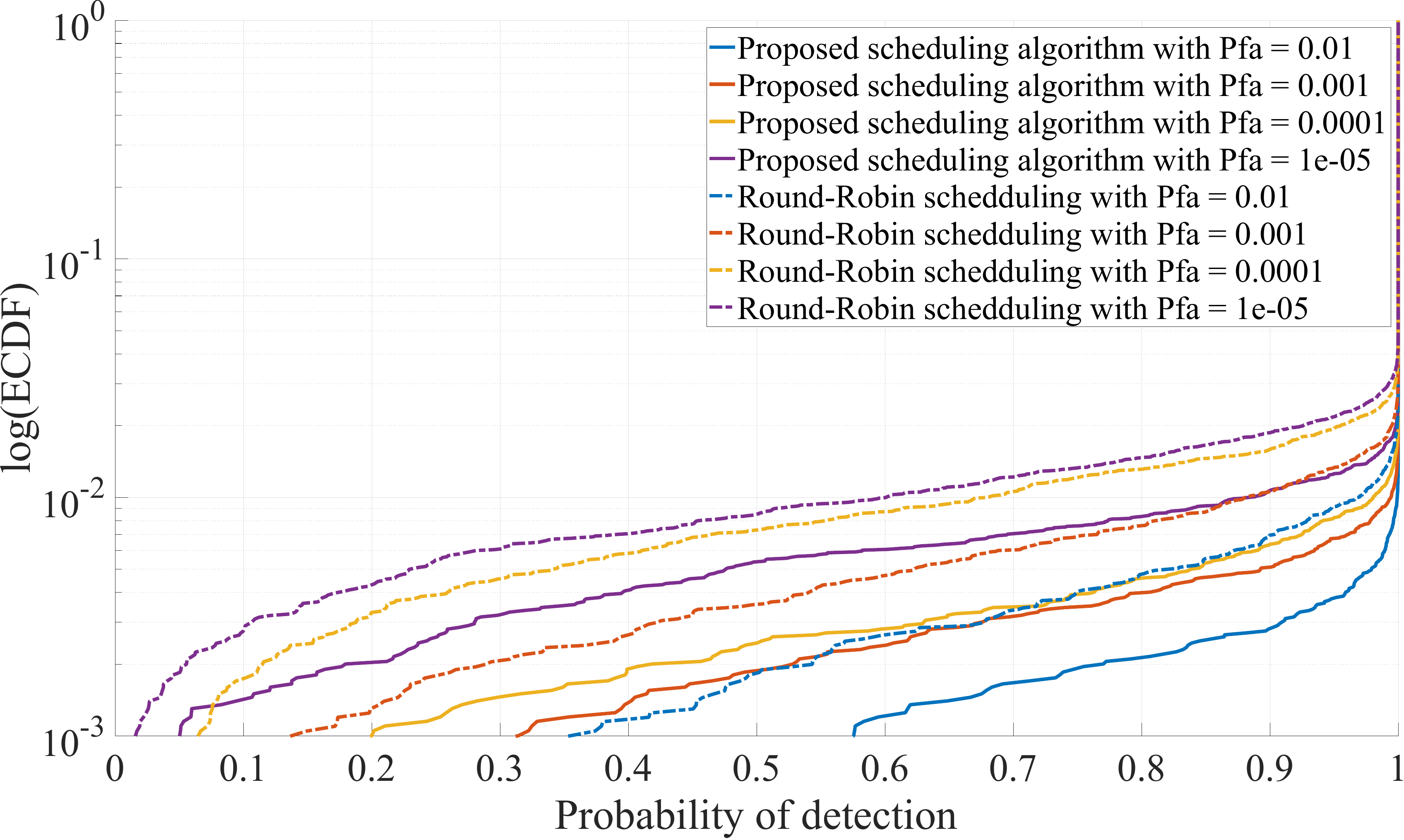}
\vspace{-1.5mm}
	\caption{Probability of detection for different false alarm rates and schedulers.}
	\label{fig:result_2}
 \vspace{-7mm}
\end{figure}

Figure \ref{fig:result_2} shows the CDF of the probability of detection for different values of the probability of false alarm, $P_{fa}$. Our proposed algorithm consistently achieves a higher probability of detection compared to the baseline. This is because the scheduler is environment-aware and dynamically allocates more resources to sensing when detection errors increase. This adaptability is especially valuable in environments where AGVs move through areas with higher clutter density, which can severely affect sensing accuracy. 

\begin{figure}[t]
\centering
\includegraphics[width=0.45\textwidth,trim = 0mm 0mm 0mm 0mm,clip]{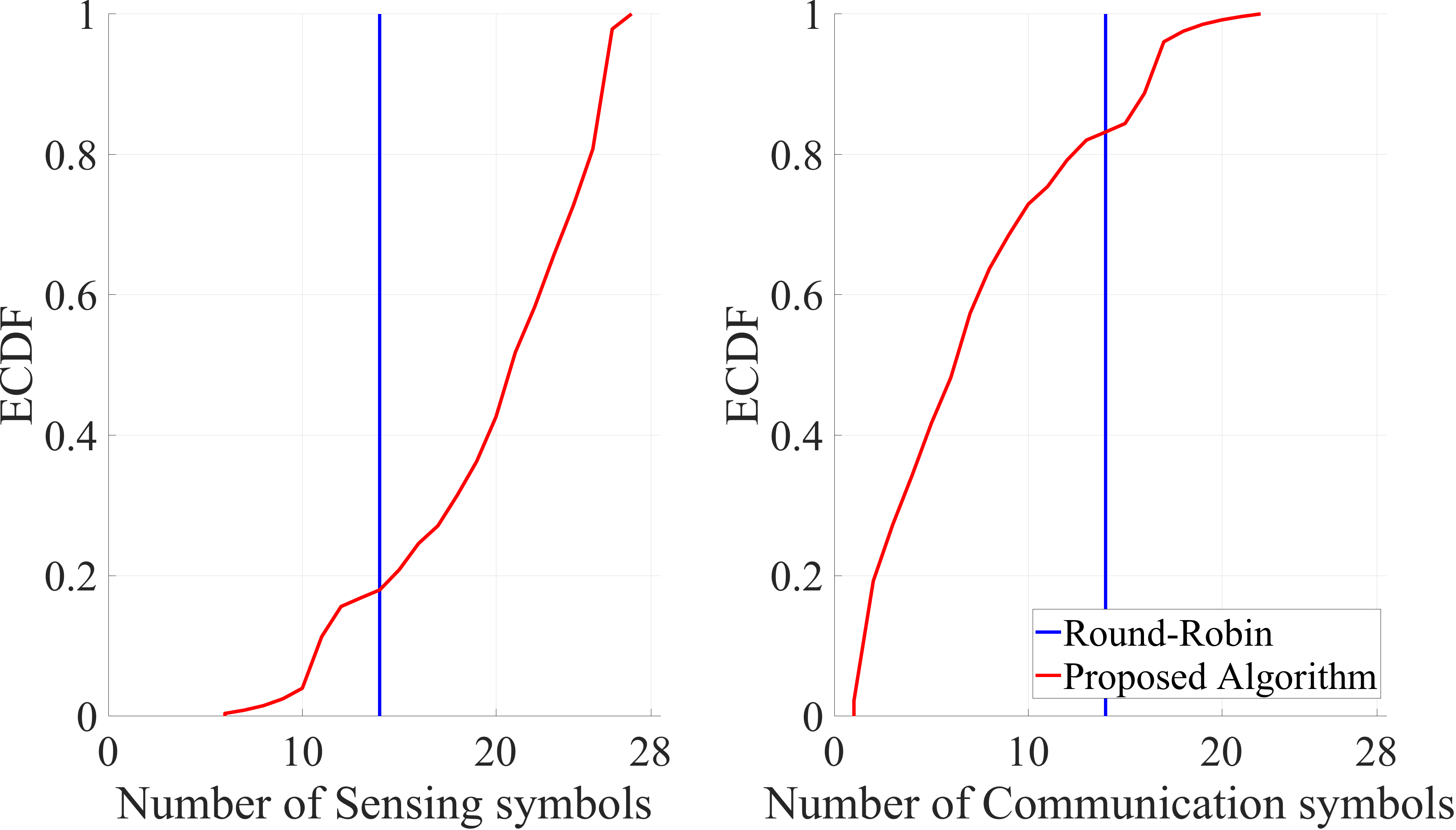}
	\caption{Comparison of the number of dedicated resources for sensing and communication.}
	\label{fig:result_3}
 \vspace{-6mm}
\end{figure}

Figure \ref{fig:result_3} illustrates the distribution of allocated resources for both sensing and communication over the entire simulation period. We observe that the proposed scheduling algorithm adaptively allocates resources between communication and sensing based on real-time demands. In particular, our proposed algorithm allocates more resources to sensing, especially in response to the high demand caused by the cluttered nature of the environment. This is necessary to maintain accurate detection, as the dense clutter can significantly impact sensing performance. At the same time, the allocated resources to communication are sufficient to effectively prevent any URLLC packet delays from breaching the survival time limit, as shown in Figure \ref{fig:results_1}. In contrast, the Round-Robin approach statically assigns resources by equally splitting the available 14 OFDM symbols to URLLC traffic, and a separate block of 14 symbols is reserved for sensing tasks.

In summary, the Nash Bargaining approach for resource allocation between sensing and communication proves to be highly effective. It successfully balances resource usage without compromising either sensing performance or communication efficiency. When the AGV traverses through a high cluttered environment, the algorithm dynamically allocates more resources to sensing to improve the probability of detection. Conversely, when the URLLC traffic load increases, the approach shifts resources toward communication to ensure that both sensing and communication demands are met. 
\section{Conclusion}
In this paper, we presented an environment-aware adaptive time-allocation scheme for ISAC systems in industrial scenarios, optimized by using the Nash Bargaining solution. Our approach effectively balances the sensing time dedicated to AGV detection and URLLC communication tasks. This ensures high detection probabilities while maintains stringent QoS requirement for URLLC traffic. Through link-level simulations, we show that our approach ensures adaptivity to changing channel and traffic conditions and outperforms traditional scheduling. As part of future work, we plan to compare our approach with machine learning-based resource allocation methods to explore the potential of further enhancing the system’s adaptability and performance in dynamic environments.
\section*{Acknowledgment}
\small
\vspace{-0.06cm}
This work was partially funded by the Federal Ministry of Education and Research Germany within the project “Open6GHub” under grant 16KISK012 and partially funded under the Excellence Strategy of the Federal Government and the Länder under grant "RWTH-startupPD 441-23".
\vspace{-0.06cm}
\bibliographystyle{IEEEtran}
\bibliography{icc_radcom}

\begin{thebibliography}{10}
\providecommand{\url}[1]{#1}
\csname url@samestyle\endcsname
\providecommand{\newblock}{\relax}
\providecommand{\bibinfo}[2]{#2}
\providecommand{\BIBentrySTDinterwordspacing}{\spaceskip=0pt\relax}
\providecommand{\BIBentryALTinterwordstretchfactor}{4}
\providecommand{\BIBentryALTinterwordspacing}{\spaceskip=\fontdimen2\font plus
\BIBentryALTinterwordstretchfactor\fontdimen3\font minus \fontdimen4\font\relax}
\providecommand{\BIBforeignlanguage}[2]{{%
\expandafter\ifx\csname l@#1\endcsname\relax
\typeout{** WARNING: IEEEtran.bst: No hyphenation pattern has been}%
\typeout{** loaded for the language `#1'. Using the pattern for}%
\typeout{** the default language instead.}%
\else
\language=\csname l@#1\endcsname
\fi
#2}}
\providecommand{\BIBdecl}{\relax}
\BIBdecl

\bibitem{Rodrigo_mag}
R.~Hernangómez \emph{et~al.}, ``{Toward an AI-Enabled Connected Industry: AGV Communication and Sensor Measurement Datasets},'' \emph{IEEE Communications Magazine}, vol.~62, no.~4, pp. 90--95, 2024.

\bibitem{factory_positioning2023}
K.~Muthineni, A.~Artemenko, J.~Vidal, and M.~Nájar, ``{A Survey of 5G-Based Positioning for Industry 4.0: State of the Art and Enhanced Techniques},'' in \emph{Proc. EuCNC/6G Summit}, 2023, pp. 120--125.

\bibitem{3GPP3}
\emph{{Feasibility Study on Integrated Sensing and Communication (Release 19), 3GPP, TR 22.837, v1.0.0, 2023.}}

\bibitem{MoeZin_icassp}
G.~Kwon, Z.~Liu, A.~Conti, H.~Park, and M.~Z. Win, ``{Integrated Localization and Communication in 3GPP Industrial Environments},'' in \emph{Proc. IEEE ICASSP}, 2024, pp. 9191--9195.

\bibitem{Silvio_servey}
S.~Mandelli \emph{et~al.}, ``Survey on integrated sensing and communication performance modeling and use cases feasibility,'' in \emph{Proc. IEEE 6GNet}, 2023, pp. 1--8.

\bibitem{navid_lnet2024}
N.~Keshtiarast \emph{et~al.}, ``{Wireless MAC Protocol Synthesis and Optimization With Multi-Agent Distributed Reinforcement Learning},'' \emph{IEEE Networking Letters}, vol.~6, no.~4, pp. 242--246, 2024.

\bibitem{ding_mobicom22}
C.~Ding \emph{et~al.}, ``{Joint precoding for MIMO radar and URLLC in ISAC systems},'' in \emph{Proc. ACM MobiCom Workshop ISACom '22}, New York, NY, USA, 2022, p. 12–18.

\bibitem{vacarubio2023}
\BIBentryALTinterwordspacing
C.~J. Vaca-Rubio \emph{et~al.}, ``Proximal policy optimization for integrated sensing and communication in mmwave systems,'' 2023. [Online]. Available: \url{https://arxiv.org/abs/2306.15429}
\BIBentrySTDinterwordspacing

\bibitem{Zinat_wcnc2024}
Z.~Behdad \emph{et~al.}, ``{Interplay Between Sensing and Communication in Cell-Free Massive MIMO with URLLC Users},'' in \emph{Proc. IEEE WCNC}, 2024, pp. 1--6.

\bibitem{ramos2024}
A.~Ramos \emph{et~al.}, ``{Enhancing Sensing-Assisted Communications in Cluttered Indoor Environments through Background Subtraction},'' \emph{arXiv preprint arXiv:2401.05763}, 2024.

\bibitem{navid_wcnc2024}
N.~Keshtiarast \emph{et~al.}, ``{Modeling and Performance Analysis of CSMA-Based JCAS Networks},'' in \emph{Proc. IEEE WCNC}, 2024, pp. 01--06.

\bibitem{nash1950bargaining}
J.~F. Nash, ``The bargaining problem,'' \emph{Econometrica}, vol.~18, no.~2, pp. 155--162, 1950.

\bibitem{touti_globe}
C.~Touati, E.~Altman, and J.~Galtier, ``{Fair power and transmission rate control in wireless networks},'' in \emph{Proc. IEEE GLOBECOM}, vol.~2, 2002, pp. 1229--1233 vol.2.

\bibitem{broadband_bargain}
H.~Yaiche \emph{et~al.}, ``A game theoretic framework for bandwidth allocation and pricing in broadband networks,'' \emph{IEEE/ACM Trans. Netw.}, vol.~8, no.~5, pp. 667--678, 2000.

\bibitem{bishoyi2024}
P.~K. Bishoyi and S.~Misra, ``{Towards Energy-And Cost-Efficient Sustainable MEC-Assisted Healthcare Systems},'' \emph{IEEE Trans. Sustain. Comput.}, vol.~7, no.~4, pp. 958--969, 2022.

\bibitem{3GPP2}
\emph{{Study on channel model for frequencies from 0.5 to 100 GHz, 3GPP, TR 38.901, v17.0.0, 2022.}}

\bibitem{Salah_jsac}
S.~E. Elayoubi \emph{et~al.}, ``{Radio Resource Allocation and Retransmission Schemes for URLLC Over 5G Networks},'' \emph{IEEE J. Sel. Areas Commun.}, vol.~37, no.~4, pp. 896--904, 2019.

\bibitem{3GPP1}
\emph{{3GPP TS 38.214, TSG RAN; NR; Physical layer procedures for data (Release 16), v16.0.0, Dec. 2019}}.

\end{thebibliography}
\end{document}